\documentclass[aps,prb,floatfix,amsmath,amssymb,preprint,eqsecnum,nofootinbib,superscriptaddress]{revtex4}
\usepackage{graphicx}
\usepackage{dcolumn}
\usepackage{bm}
\usepackage{epstopdf}
\usepackage{setspace}   

\newcommand{\beq}{\begin{equation}}
\newcommand{\eeq}{\end{equation}}
\newcommand{\ba}{\begin{array}{ccc}}
\newcommand{\ea}{\end{array}}
\newcommand{\nn}{\nonumber}
 \renewcommand{\d}{\partial}
\def\bea{\begin{eqnarray}}
\def\eea{\end{eqnarray}}

\def\<{\langle}
\def\>{\rangle}

\usepackage{amsmath}
\usepackage{amssymb}

\begin{document}

\title{Bosonic topological insulator in three dimensions and the statistical Witten effect}

\author{Max A. Metlitski}
\affiliation{Kavli Institute for Theoretical Physics, UC Santa Barbara, CA 93106}
\author{C. L. Kane}
\affiliation{Department of Physics and Astronomy, University of Pennsylvania, Philadelphia, PA 19104}
\author{Matthew P. A. Fisher}
\affiliation{Physics Department, UC Santa Barbara, CA  93106}

\date{\today\\
\vspace{1.6in}}
\begin{abstract}
It is well-known that one signature of the three-dimensional electron topological insulator is the Witten effect: if the system
is coupled to a compact electromagnetic gauge field, a monopole in the bulk acquires a half-odd-integer polarization charge.
In the present work, we propose a corresponding phenomenon for the topological insulator of bosons in 3d protected by particle number conservation
and time-reversal symmetry. We claim that although a monopole inside a topological insulator of bosons can remain electrically neutral, 
its statistics are transmuted from bosonic to fermionic. 
We demonstrate that this ``statistical Witten effect" directly implies that if the surface of the topological
insulator is neither gapless, nor spontaneously breaks the symmetry, it necessarily supports an intrinsic two-dimensional topological order.
Moreover, the surface properties cannot be fully realized in a purely
2d system. We also confirm that the surface phases of the bosonic topological insulator proposed by Vishwanath and Senthil (arXiv:1209.3058) provide
a consistent termination of a bulk exhibiting the statistical Witten effect. In a companion paper, we will provide an explicit field-theoretic, lattice-regularized,
construction of the 3d topological insulator of bosons, employing a
parton decomposition and subsequent condensation of parton-monopole composites.


%

\end{abstract}

\maketitle

\section{Introduction}
The discovery of electronic topological insulators\cite{TI,FuKaneMele,MooreBalents,Roy,Hasan} has fueled a growing interest in so-called symmetry-protected topological (SPT) phases of matter.
These are phases with a fully gapped bulk spectrum, whose stability, as their name suggests, is guaranteed by a global symmetry.
When the symmetry is present, one cannot continuously deform a non-trivial SPT state to a trivial product state without a bulk phase transition.
On the other hand, when the symmetry is absent, an SPT phase may be continuously connected to a trivial product state. Thus, we may say that 
SPT phases have no ``intrinsic" topological order: they have a unique ground state on any closed manifold and possess no fractional bulk excitations.
Although their bulk spectrum is trivial, SPT phases display highly unusual edge properties. Namely, in one  dimension, the zero-dimensional edge is gapless,
in two dimensions, the one-dimensional edge is either gapless or spontaneously breaks a symmetry, finally, in three dimensions, the
two-dimensional edge is either gapless, spontaneously breaks a symmetry or carries intrinsic topological order. Moreover, in all these cases, the properties
of the edge of a $d$-dimensional non-trivial SPT phase cannot be realized in a purely $d-1$ dimensional system (at least if the global symmetry in this lower dimensional system
acts in a conventional way).

An example that illustrates the above properties is the three-dimensional electronic topological insulator, which is a phase of matter protected by the symmetry group $U(1)\ltimes Z^T_2$, with $U(1)$ - the 
charge conservation symmetry and $Z^T_2$ - time reversal.\footnote{The semi-direct product $\ltimes$ means that the anti-unitary time reversal operator ${\cal T}$ and the $U(1)$ rotations $g$ do not commute,
${\cal T}^{-1} g {\cal T} = g^{-1}$. Moreover, time-reversal is realized projectively here, ${\cal T}^2 = -1$. Note that this corresponds to the standard transformation properties of spinfull fermions $f_\alpha$ under time-reversal: ${\cal T}:\,f_\alpha \to \epsilon_{\alpha \beta} f_\beta$.} The bulk spectrum is gapped and trivial, however, the
surface supports an odd number of gapless Dirac cones and preserves the full $U(1)\ltimes Z^T_2$ symmetry. It is well known (and will be reviewed below) that such properties cannot be realized in a purely two-dimensional system.\cite{Redlich1,Redlich2,FionaMike} Moreover, one can distinguish a topological insulator from a trivial one even in the bulk by considering its response to an external {\it compact} electromagnetic field. The topological insulator exhibits electromagnetic response with a topological $\theta$ angle, $\theta = \pi$, while the trivial insulator has $\theta = 0$. The value of the $\theta$ angle is manifested in the Witten effect\cite{Witten}: if one inserts a magnetic monopole into a topological insulator, it acquires a half-odd-integer charge.\cite{SCZ,Franz}

Over the past few years remarkable progress has been made in understanding SPT phases. In particular, all symmetry protected phases of non-interacting fermions in arbitrary dimension have been classified.\cite{Ludwig, KitaevFree} Likewise,
it is believed that all possible interacting gapped phases of bosons and fermions in one dimension have been constructed.\cite{Fidkowski1d,Ari,Wen1d,Cirac,Wen1dfull} The phases of bosons in $d=1$ are classified by the set of projective representations of the symmetry group $G$, 
which is identical to the second cohomology group of $G$.\cite{Wen1d, Wen1dfull}  A Jordan-Wigner transformation can be used to extend this result to one-dimensional fermionic systems.
A generalization of the one-dimensional cohomology classification to higher-dimensional bosonic systems has been recently proposed.
\cite{WenCohoBoson} However, the physical properties of the resulting higher dimensional SPT phases are just starting to be exposed. 
In the case of two dimensions, it has been shown\cite{Ashvin2d} that a $K$-matrix construction - a multi-component Abelian Chern-Simons theory - analogous to the one used to 
describe topologically ordered Abelian quantum Hall states, reproduces the cohomology classification of SPT phases and gives a direct description of their gapless edge modes.

In the case of three dimensions, an important advance has been made by Vishwanath and Senthil (VS), \cite{VS} who have suggested effective bulk and surface field theories for a number 
of bosonic SPT phases. In particular, VS have proposed an effective theory for the bosonic SPT phase with the symmetry group $U(1) \ltimes Z^T_2$, which is the direct bosonic analogue of the
three-dimensional fermionic topological insulator.\footnote{In the bosonic case, time-reversal is realized non-projectively ${\cal T}^2 = 1$, and boson operators transform simply as, ${\cal T}:\,b \to b$.} In the present paper we will focus on this particular phase and refer to it simply as the bosonic topological insulator. 

VS have claimed that the bosonic topological insulator is characterized by an electromagnetic response with the $\theta$ angle, $\theta = 2\pi$. This was interpreted to mean that one possible way to terminate the bulk topological insulator at the surface is to have a fully gapped surface state with no intrinsic topological order, which spontaneously breaks the time-reversal symmetry and has a Hall conductivity $\sigma_{xy} = \pm 1$. A general physical argument of Ref.~\onlinecite{LevinSenthil}, complemented by the explicit $K$-matrix construction of Ref.~\onlinecite{Ashvin2d}, indicates that purely two-dimensional bosonic SPT phases with a global $U(1)$ symmetry always have an even integer $\sigma_{xy}$, so the surface proposed by VS cannot be realized in a strictly two-dimensional model.   Moreover, one can go between the time-reversal conjugate surface states with $\sigma_{xy} = 1$ and $\sigma_{xy}= -1$ by ``painting" a layer of the 2d bosonic insulator with $\sigma_{xy} = 2$ on the surface, which is consistent with time-reversal symmetry being broken only on the surface.

By driving surface phase transitions, VS were then able to construct alternative surface phases of the bosonic topological insulator, where the time-reversal symmetry is restored. These include:

i) a gapped state which preserves the full $U(1)\ltimes Z^T_2$ symmetry, but carries an intrinsic $Z_2$ topological order. This state has the unusual property that both the electric and magnetic anyons carry charge $1/2$ under the global $U(1)$ symmetry.

ii)  a superfluid state, which spontaneously breaks the global $U(1)$ symmetry, but leaves the $Z^T_2$ symmetry intact. This state is characterized by unusual vortex properties that will be discussed in more detail below.

iii) certain (multi)-critical (gapless) states preserving both the $U(1)$ and the $Z^T_2$ symmetry, which will not be discussed in the present paper.  

The work of VS has left three questions open:

I. How to distinguish a non-trivial bosonic topological insulator from a trivial one in the bulk? In the case of the fermionic topological insulator with $\theta = \pi$, the Witten effect provided such a bulk signature. However, if we accept that bosonic topological insulators have $\theta = 2 \pi$, the charge carried by a magnetic monopole inside the topological insulator is quantized in integer units, just as in a trivial insulator with $\theta = 0$. In particular, in both cases monopoles can carry zero electric charge.

II. What is the fundamental physical reason why the surface of a 3d bosonic topological insulator, independent of what phase it is in,  cannot be realized in a purely 2d system?
For the case of the fermionic topological insulator, such a reason exists: a putative 2d model exhibiting the same properties as the surface cannot be consistently coupled to a {\it compact} electromagnetic gauge field.\cite{Redlich1,Redlich2,FionaMike}

III. Why is the surface of a bosonic topological insulator necessarily either gapless, symmetry broken or topologically ordered?

In the present paper we resolve the above three questions and demonstrate that they are intimately linked. We show that one can distinguish between the trivial and non-trivial bosonic topological insulators in the bulk by coupling them to a {\it weakly fluctuating compact} electromagnetic field. It is, indeed, appropriate to think of the trivial and non-trivial bosonic topological insulators as having $\theta = 0$ and $\theta = 2\pi$. Although magnetic monopoles in both insulators can be electrically neutral, their statistics are different. In the trivial insulator ($\theta = 0$), monopoles  are bosons, while in the non-trivial insulator ($\theta = 2\pi$), monopoles are fermions. We call this phenomenon the ``statistical Witten effect." Once we dial $\theta$ to $4 \pi$, monopoles are again bosons. Hence, the $\theta$ angle in a bosonic insulator is periodic modulo $4 \pi$. This is in sharp contrast to a fermionic insulator, where monopoles have bosonic statistics both at $\theta = 0$ and at $\theta = 2\pi$, so the $\theta$ variable is periodic modulo $2 \pi$ in accordance to the common lore.

We note that the idea of distinguishing different SPT phases by ``gauging" the global symmetry group has been utilized before. An example is provided by phases with a global $Z_2$ symmetry in two dimensions. In this case, the cohomology classification gives one non-trivial SPT phase. As shown in Ref.~\onlinecite{LevinGu}, if one starts with the trivial phase and couples it to a weakly fluctuating $Z_2$ gauge field, one obtains a system with a ``toric code" topological order. The $Z_2$ fluxes (visons) in this state have bosonic statistics. On the other hand, if one starts with the non-trivial SPT phase and gauges the $Z_2$ symmetry, one obtains a system with a ``double-semion" topological order. The fluxes in this state have semionic statistics. The statistical Witten effect in a three-dimensional topological insulator is a direct analogue of this phenomenon. We would like to mention that a general duality between SPT phases and ``weakly fluctuating" gauge theories was discussed in Ref.~\onlinecite{Wenduality}. It is believed that the duality transformation can be physically interpreted as gauging the global symmetry of the SPT.

Returning to the 3d bosonic topological insulator, the statistical Witten effect in the bulk gives a clue as to why the surface physics cannot be fully realized in a 2d system.  It turns out that, as in the fermionic case, a putative 2d model exhibiting the same properties as the surface of a bosonic topological insulator cannot be consistently coupled to a two-dimensional compact electromagnetic gauge field. Such a coupled theory would have instanton events in space-time, where the magnetic flux through the 2d surface changes by $2\pi$. We will show that these events are accompanied by the creation of a single fermion excitation out of the 2d vacuum. This is impossible in a local theory. Hence, the surface cannot exist without the bulk! When the bulk is present, the above ``anomaly" of the surface theory is cured as follows. In the three-dimensional world, the instanton tunneling event of the 2d surface theory corresponds to the passage of a magnetic monopole from the trivial vacuum outside into the topological insulator. In the process, the monopole changes its statistics from bosonic to fermionic and {\it leaves another fermion on the surface}.\footnote{This picture is slightly simplified; we will give a more general discussion below.} Thus, fermions are created in pairs and no violation of locality occurs. 

Besides demonstrating that the surface of a bosonic topological insulator cannot exist purely in 2d, we will use the statistical Witten effect to argue that if the surface state is neither gapless,
nor spontaneously breaks the global symmetry it {\it must} possess intrinsic  topological order.


We now give a more detailed exposition of the above arguments. In this paper, our discussion of the bulk and surface properties of the bosonic topological insulator will
be guided only by general considerations of symmetry and locality, without reference to a construction of this phase. In a companion paper,\cite{Forthcoming} we will supplement
the present conclusions by an explicit field-theoretic construction of the bosonic topological insulator, which displays the statistical Witten effect in the bulk and realizes
the phases proposed by VS at its surface.

This paper is organized as follows. In section \ref{sec:Witt}, we briefly reveiw the $\theta$-parameter, the Witten effect and its role in fermionic topological insulators. Section \ref{sec:statWitt}
is devoted to the statistical Witten effect in the bulk of a bosonic topological insulator. Section \ref{sec:surf} discusses the link between the statistical Witten effect in the bulk and the 
physics at the surface of the bosonic topological insulator. Section \ref{sec:coho} discusses a potential place of the bosonic topological insulator phase, to which this paper is devoted,
within the general cohomology classification of Ref.~\onlinecite{WenCohoBoson}.

\section{The Witten effect and fermionic topological insulators}
\label{sec:Witt}

We begin with a brief review of the $\theta$-parameter, the Witten effect and its role in fermionic topological insulators.

Consider a fully gapped fermion insulator with no intrinsic topological order, i.e. with a unique ground state on any closed manifold and with excitations carrying an integer electric charge. It is useful to study the response of the insulator to an external electromagnetic field, $A_{\mu}$. In 3d, in addition to the standard Maxwell term describing the electric polarizability/magnetic permeability of an insulator, the electromagnetic response involves the topological $\theta$-term,
\beq S_\theta = \frac{i \theta}{32 \pi^2}  \int d^3x d\tau \epsilon_{\mu \nu \lambda \sigma} F_{\mu \nu} F_{\lambda \sigma} \label{eq:Stheta}\eeq
where $F_{\mu \nu} = \d_{\mu} A_{\nu} - \d_{\nu} A_{\mu}$. For smooth configurations of the electromagnetic field on a four-dimensional space-time torus, the imaginary time action $S_\theta$ evaluates to $S_\theta = i \theta n$, with integer $n$. In this sense, the $\theta$-parameter is periodic modulo $2 \pi$. Moreover, under time reversal, $\theta \to -\theta$. Hence, naively, the distinct time-reversal invariant values of $\theta$ are $\theta = 0$ and $\theta = \pi$. Below, we will see that this conclusion is correct for fermionic insulators, but not for bosonic ones. It turns out that both $\theta = 0$ and $\theta = \pi$, can, indeed, be realized already in non-interacting time-reversal invariant fermionic systems. Trivial non-interacting fermionic insulators have $\theta = 0$. On the other hand, non-interacting fermionic topological insulators have $\theta = \pi$.\cite{SCZ} This provides a formal distinction between the two classes of fermionic insulators with time-reversal symmetry which does not rely on surface properties.

Another bulk manifestation of the $\theta$ term is the so-called Witten effect.\cite{Witten} Since we are discussing lattice systems here, the insulator with a global $U(1)$ symmetry can be consistently coupled to a {\it compact} electromagnetic gauge field. Hence, monopole configurations of the magnetic field are allowed. As shown by Witten, in the presence of the $\theta$-angle, a magnetic monopole with flux $2\pi m$ also carries an electric charge $q = n + \frac{\theta m}{2\pi}$, with integer $n$.\footnote{We would like to stress that in the presence of monopole configurations,  $S_\theta$, Eq.~(\ref{eq:Stheta}), like the rest of the terms in the continuum field theory needs to be regularized. The Witten effect, however, is independent of this regularization.}  The integer $n$ corresponds to the freedom of adding extra particle excitations of the insulator on top of the monopole. We note that the allowed combinations of electric charge $q$ and magnetic charge $m$ are invariant under $\theta \to \theta + 2\pi$, reaffirming the interpretation of $\theta$ as a periodic variable. Hence, if a single magnetic monopole is placed inside a fermionic topological insulator, it acquires a half-odd-integer polarization charge.\cite{SCZ,Franz} Thus, the Witten effect gives a clear way to distinguish trivial and topological fermionic insulators. 

The Witten effect is directly connected to the reason why the surface of a 3d fermionic topological insulator cannot be realized in a two-dimensional model. Imagine an interface between a topological insulator and the vacuum (or a trivial insulator). Consider taking a magnetic monopole in vacuum and dragging it across the surface with the non-trivial insulator. The monopole in vacuum carries no charge but acquires a half-odd-integer charge once inside the topological insulator. Since electric charge is conserved, this means that a half-odd-integer charge must be left on the surface. 

Now suppose the surface of the topological insulator {\it could} be realized purely in a 2d lattice model. Then we can couple this 2d model to a compact $U(1)$ gauge field. Imagine starting with a configuration
with no magnetic flux through the surface. This configuration is equivalent to one where flux $2 \pi$ passes through a single plaquette of the surface. Now let this flux $2 \pi$ slowly expand to form a smooth
flux distribution over some portion of the surface.  
This process is an instanton event in space-time, where the flux changes from $0$ to $2 \pi$. Now, from the point of view of the 3d system, the instanton event corresponds to the passage of a monopole through the surface of the topological insulator. As noted above, this process must deposit a half-odd-integer charge on the boundary. Thus, in a pure 2d theory of the boundary, local instanton events violate charge conservation. The boundary theory taken by itself is, therefore, inconsistent. We may say that the $U(1)$ symmetry on the surface is ``anomalous." In the three-dimensional world, this surface anomaly is compensated by the Witten effect in the bulk and so charge is conserved.

The above argument indicates that the surface of the topological insulator exhibits a particular kind of anomaly, independent of what phase the surface is in. Let us now demonstrate the anomaly for a particular realization of the surface. The most commonly discussed surface phase is a time-reversal invariant state with a single gapless 2d Dirac cone. Although it is possible to show that a putative 2d system with such properties cannot be consistently coupled to a 2d compact $U(1)$ gauge-field,\cite{Redlich1, Redlich2, FionaMike}  the demonstration is much simpler for a different realization of the surface. Imagine spontaneously (or explicitely) breaking the $Z^T_2$ symmetry on the surface. This gaps out the Dirac cone and results in the surface having a Hall conductivity $\sigma_{xy} = \pm 1/2$, i.e. a background gauge field $A_{\mu}$ induces an electromagnetic ``polarization" current on the surface,
\beq J^{pol}_{\mu} = \frac{\sigma_{xy}}{2 \pi} \epsilon_{\mu \nu \lambda} \d_{\nu} A_{\lambda}  . \label{eq:JHall}\eeq
 We focus on one of the time-reversal conjugate states, say the one with $\sigma_{xy} = 1/2$. The only excitations of this state are gapped fermions with charge $1$. We now show that such a state cannot exist purely in 2d.

Imagine first coupling the putative 2d model to a non-compact $2+1$ dimensional $U(1)$ gauge field $A_{\mu}$. Upon integrating the gapped fermions out, we obtain an effective Chern-Simons action for the electromagnetic field,
\beq S = \frac{i k}{4 \pi} \int d^2 x d \tau \epsilon_{\mu \nu \lambda} A_{\mu} \d_{\nu} A_{\lambda}   , \label{eq:CS}\eeq 
with $k = \sigma_{xy} = 1/2$. If the system is realized in a 2d lattice model, we can also couple it to a compact $U(1)$ gauge field, so we should be able to promote $A_{\mu}$ in Eq.~(\ref{eq:CS}) to a compact field. It is well-known,\cite{Teitelboim} that the level $k$ of the Chern-Simons action (\ref{eq:CS}) must be an integer if the $U(1)$ gauge field is compact and the charged excitations are fully gapped and have integer charge, so a level $k = 1/2$ is inconsistent. We now sketch the argument for this.

Imagine a gauge field configuration with no magnetic flux through the surface and zero electric charge. As before, consider an instanton event where flux $2\pi$ is nucleated through a single plaquette and then allowed to expand to a smooth distribution. According to Eq.~(\ref{eq:JHall}), this flux distribution carries a polarization charge $1/2$. If the theory was purely two-dimensional, a local charge $-1/2$ quasiparticle excitation must also be created on the surface
during the instanton event in order to conserve the total electric charge.\footnote{This is just the standard Laughlin argument which states that a 2d system with a Hall conductivity $\sigma_{xy}$ should possess excitations with charge $q = \sigma_{xy}$.} 
However, the only excitations of the surface in this $\sigma_{xy} = 1/2$ phase are gapped fermions with charge $1$. So, a purely two-dimensional theory is not consistent: instanton events necessarily violate charge conservation by a half-odd-integer. Of course, this is precisely the property that the surface of a topological insulator should satisfy: charge $-1/2$ from the surface is transferred onto the monopole that tunnels through it. Processes where charge $-1/2-n$ is transferred onto the monopole and $n$ charge $1$ fermions are locally excited on the surface are likewise allowed.


We can use a variation of the above argument to show that if the surface of a fermionic topological insulator is neither gapless, nor spontaneously breaks the time-reversal symmetry then it is necessarily topologically ordered. 
Indeed, suppose the surface is fully gapped and preserves the time-reversal symmetry. The surface Hall conductivity must then be $0$. Now, imagine dragging a magnetic monopole through the surface of the topological insulator. Since $\sigma_{xy} = 0$, the magnetic flux through the surface induces no surface polarization charge. Hence, the half-odd-integer charge left by the monopole on the surface
must be a localized surface quasiparticle excitation. So, the surface supports excitations with a fractional charge and, therefore, by the standard flux threading argument, the ground state of the system on a 3d solid torus is degenerate. QED.

\section{The statistical Witten effect and the bosonic topological insulator}
\label{sec:statWitt}
Let us turn our attention to 3d bosonic insulators with no intrinsic topological order. Now the bulk excitations are bosons $b$ with integer charge. We may again consider the response of the system to a compact $U(1)$ gauge field: a $\theta$-term (\ref{eq:Stheta}) in the effective action will generally be induced. Our discussion of the $\theta$-term and the Witten effect in the previous section was independent of the statistics of the bulk excitations. In particular, for smooth configurations of the electromagnetic field on a four-torus, we still have $S_\theta = i \theta n$, with an integer $n$. Moreover, magnetic monopoles with flux $2 \pi m$ still carry electric charge $q = n + \frac{\theta m}{2 \pi}$, with $n$ an integer. Thus, we may naively conclude that $\theta$ is periodic modulo $2 \pi$ and that the time-reversal invariant points are $\theta = 0$ and $\theta = \pi$, corresponding to bosonic trivial and  topological insulators, respectively. It turns out that this conclusion is incorrect. In reality, in a bosonic insulator, the $\theta$-variable is periodic modulo $4\pi$ and the distinct time-reversal invariant values of $\theta$ are $\theta = 0$ (trivial insulator) and $\theta = 2\pi$ (topological insulator), while $\theta = \pi$ always breaks the time-reversal symmetry.

To reach the above conclusion we need a somewhat finer grating than the ordinary Witten effect discussed above. Namely, we need to consider the statistics of monopole excitations.

Let us, again, couple our bosonic insulator to a weakly fluctuating compact $U(1)$ gauge field. The gauge field in the bulk of the insulator will be in a Coulomb phase - i.e. there will be a gapless photon excitation, described by Maxwell electrodynamics. In addition to the gapless photon, there will be gapped bosonic excitations $b$ with integer electric charge. The photon mediates a standard Coulomb interaction $V(r) = e^2 q_1 q_2 c/(4 \pi  r)$ between two static excitations with charges $q_1$ and $q_2$. Here, $e$ is a dimensionless coupling constant and $c$ is the velocity of the photon. 

Besides the electric charge excitations, the theory will possess gapped magnetic monopole excitations with flux $2 \pi m$. For now, let us consider the trivial bosonic insulator with $\theta = 0$. The monopoles then have bosonic statistics and carry electric charge $0$. Two static monopoles with fluxes $2 \pi m_1$ and $2\pi m_2$ experience a Coulomb interaction,  $V(r) = (2 \pi)^2 m_1 m_2 c/(4 \pi e^2  r)$.

The photon also mediates a ``statistical" interaction between the charges and the monopoles. It is known\cite{Goldhaber} that this statistical interaction results in the bound state of $n$ charges and $m$ monopoles having statistics $(-1)^{nm}$, where $+1$ corresponds to bosonic statistics and $-1$ to fermionic statistics. It is common to refer to such general bound states as dyons.

\begin{figure}
\includegraphics[width=0.8\textwidth]{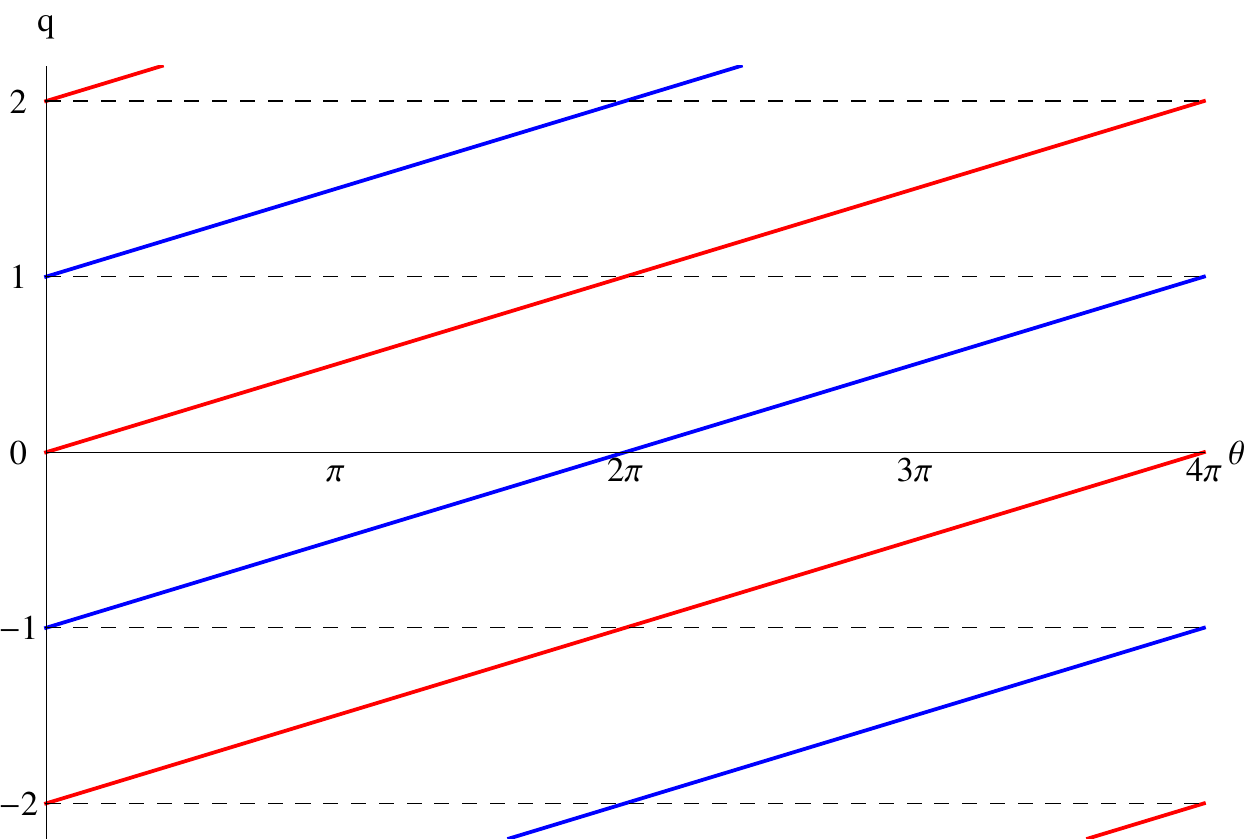}
\caption{Electric charge $q$ of dyons with magnetic flux $2 \pi$ as a function of the $\theta$-angle in a bosonic insulator. Red lines denote dyons with bosonic statistics
and blue lines - dyons with fermionic statistics. Although the allowed values of electric charge are invariant under $\theta \to \theta + 2 \pi$,  the corresponding statistics
is periodic only modulo $4\pi$.} \label{fig:Statistics}
\end{figure}

Now, let us reintroduce the $\theta$ variable. As we start tuning $\theta$ away from $0$, a monopole with flux $2 \pi m$ acquires an electric ``polarization" charge $\theta m/{2 \pi}$. Thus, a dyon which had
electric/magnetic charges $(n,m)$ at $\theta = 0$, now carries charges $(n + \theta m/2 \pi, m)$. For $\theta= 2\pi$, both the total (physically observable) electric charge,
$q = n + \theta m /2\pi$, and the magnetic charge $m$ are integer.  However, the polarization charge does not affect the dyon statistics.\cite{MacKenzie} 
This is consistent with the fact that statistics in three-dimensions are either bosonic or fermionic. Thus, when $\theta$ is finite, the statistics of a dyon with charges $(n+ \theta m/2 \pi, m)$ remain at $(-1)^{n m}$. Let us express the statistics of a dyon at a finite value of $\theta$ in terms of its total  electric charge $q = n + \theta m/2 \pi$. This yields $(-1)^{q m - \theta m^2/2 \pi}$ for the statistics of a $(q,m)$ dyon at a general $\theta$.  Thus, when $\theta = 0$, the statistics of a $(q,m)$ dyon is $(-1)^{q m}$, but when $\theta = 2 \pi$, the statistics of a $(q , m)$ dyon
($q$ and $m$ - both integer) is $(-1)^{q m + m}$. This means that although the allowed electric/magnetic charges at $\theta = 0$ and $\theta = 2 \pi$ are the same, the dyon statistics are in general different! In particular, excitations with total electric charge $0$ and magnetic flux $2\pi$ are bosons at $\theta = 0$ and fermions at $\theta = 2 \pi$. Below, we refer to these excitations simply as single monopoles. Note that once $\theta = 4 \pi$, the statistics of a $(q,m)$ dyon, again with $q$ and $m$ - both integer, return to $(-1)^{q m}$. Thus, we conclude that in a bosonic  insulator, the $\theta$ variable is periodic only modulo $4 \pi$. Therefore, the only possible distinct time-reversal invariant points are $\theta = 0$ and $\theta = 2 \pi$. In particular, the point $\theta = \pi$ manifestly breaks time-reversal invariance: here the $(1/2, 1)$ dyon has bosonic statistics, but its putative time-reversal partner $(1/2, -1)$ has fermionic statistics.

We identify the values $\theta = 0$ and $\theta = 2 \pi$ with trivial and topological insulators of bosons. The bulk signature of a bosonic topological insulator is the statistical Witten effect: monopole excitations carry fermionic statistics. Below, we will show that this identification is consistent with the surface phases of the bosonic topological insulator proposed by VS.

Before we proceed to the surface of a bosonic topological insulator, let us reconsider fermion insulators from the point of view of dyon statistics. Let us first take the trivial fermion insulators with $\theta = 0$.
Here, unit electric charges have fermionic statistics, while monopoles have bosonic statistics. The statistical interaction between charges and monopoles, therefore, endows the $(n,m)$ dyon with statistics $(-1)^{n m + n}$.
In particular, dyons with flux $2 \pi$ and arbitrary electric charge are bosons. Now, let us turn on a finite $\theta$ angle. The statistics of an excitation with total electric charge $q$ and magnetic flux $m$ is now, $(-1)^{q m + q - \theta m (m+1)/(2 \pi)}$. The above expression is invariant under $\theta \to \theta + 2\pi$. So the $\theta$ variable in a fermionic insulator is, indeed, periodic modulo $2\pi$, as is commonly assumed, and the time reversal invariant points are $\theta = 0$ and $\theta = \pi$. In particular, at $\theta = \pi$, the $(1/2, \pm 1)$ dyons are both bosons, while $(0, \pm 2)$ dyons are both fermions.

Before we conclude this section, we briefly mention a more formal field-theoretic way to deduce the periodicity of the $\theta$ angle in bosonic and fermionic systems, which considers only smooth configurations of the electromagnetic field without monopole defects.\cite{Witten4d,witten} Here, one places the system on a generic closed four-manifold, instead of the four-torus, which we considered so far. It turns out that on a most general four-manifold, $S_\theta = i \theta n$, with $n$\,- a {\it half}-integer. An example of a manifold where a gauge field configuration with half-odd $n$ exists is $\bf{C} \bf{P}^2$, which allows for $n = 1/2$. Thus, the $\theta$ variable is periodic at least modulo $4\pi$. However, theories with fermion degrees of freedom require that the space-time manifold be endowed with a spin structure. Not all manifolds admit a spin structure; those that do, have integer $n$. Thus, in theories with fermions, $\theta$ is quantized modulo $2\pi$. 

\section{Surface of the bosonic topological insulator and the statistical anomaly}
\label{sec:surf}
The statistical Witten effect displayed by the bulk of a bosonic topological insulator strongly constrains the properties of its surface. Indeed, as usual, let's couple the system to a weakly fluctuating compact $U(1)$ gauge field.
As we drag a monopole through the surface, it changes its statistics from bosonic to fermionic. Since in a local theory fermions are created in pairs, this means that a fermion must be left on the surface. 
Thus, from the point of view of the 2d surface, instanton events create single fermion excitations, (apparently) violating locality. Therefore, the surface cannot be realized in a local two-dimensional model. 
We will call this property of the surface, ``the statistical anomaly." We note that this ``anomaly," must be manifested by the surface,
independent of what phase it is in, in order to have a consistent three-dimensional theory.

It turns out that the discussion above is slightly simplified. In reality, the weakly fluctuating $U(1)$ gauge theory in the bulk is in the Coulomb phase and long-range statistical interactions 
between charged excitations on the surface and monopoles in the bulk are present. As we will see in an explicit example below, this means that in the full 3d theory, the point excitation left by the monopole 
on the surface need not be a fermion. Nevertheless, a putative 2d model with the properties of the surface must still exhibit the statistical anomaly, when coupled to a 2d compact $U(1)$ gauge field.

We now turn to the various possible surface phases of a bosonic topological insulator proposed by VS. We will show that all of these states exhibit the statistical anomaly in 2d and discuss how the statistical Witten effect
resolves the anomaly in the full 3d theory. We will then present an argument showing that if the surface of a bosonic topological insulator is neither gapless, nor spontaneously breaks the symmetry it must carry topological order.



\subsection{Surface phases with $\sigma_{xy} = \pm 1$.}

The simplest surface phase proposed by VS is an insulator with no intrinsic topological order and a Hall conductivity $\sigma_{xy} = \pm 1$. This state spontaneously breaks the time-reversal symmetry.
Its excitations are gapped bosons $b$ with integer charge. Let us show that a purely 2d system with such properties would exhibit a statistical anomaly and so cannot be realized.\footnote{Our argument 
is very similar to one given in Ref.~\onlinecite{LevinSenthil}. The only difference is that we treat the electromagnetic field as a weakly fluctuating dynamical field, instead of an external field.} Imagine first coupling
the putative 2d model to a weakly fluctuating 2d non-compact $U(1)$ gauge field $A_{\mu}$. We again 
obtain a 2d Chern-Simons effective action but now with level $k = 1$,
\beq S = \int d^2 x d \tau \left(\frac{1}{4e^2} F^2_{\mu \nu} + \frac{i k}{4 \pi} \epsilon_{\mu \nu \lambda} A_{\mu} \d_{\nu} A_{\lambda}   + i A_{\mu} J^b_{\mu} \right)\label{eq:CS2}\eeq 
Here, in addition to the Chern-Simons term, we've kept the Maxwell term $F^2_{\mu \nu}$ in the action, as well as an explicit coupling of the gauge field to the gapped boson excitations, whose current is denoted by $J^b_{\mu}$. Thus, the total electric current is given by the sum of the polarization current (\ref{eq:JHall}) and the boson current $J^b_{\mu}$,
\beq J^{EM}_{\mu} = \frac{k}{2 \pi} \epsilon_{\mu \nu \lambda} \d_{\nu} A_{\lambda} + J^b_{\mu}\eeq

As is well known, the 2+1 dimensional theory (\ref{eq:CS2}) is fully gapped.
The effect of the Chern-Simons term with $k=1$ is to dynamically attach magnetic flux $-2 \pi$ to the $b$ excitations, which have intrinsic electric charge $1$ as seen in (\ref{eq:CS2}). 
As a result, the statistics of $b$ is transmuted from bosonic to fermionic. Moreover, since the magnetic flux $\Phi$ carries an electric polarization charge $\Phi/2\pi$, the $b$ particle with flux $-2\pi$ attached
is electrically neutral.

Now, if the phase under consideration can be realized in a 2d lattice model, we should be able to promote $A_{\mu}$ to a compact gauge field. Imagine an instanton event in the putative 2d theory (\ref{eq:CS2}). Start with a configuration with no flux and no charge, nucleate flux $2\pi$ through a single plaquette and let this flux expand into a 
continuous distribution. The flux will then carry a polarization charge $1$. Therefore, in order for the total charge to be conserved, a local charge $-1$ quasiparticle
excitation must also be created during the instanton event. This excitation must be identified with the $b$ (anti)-particle. Thus, an instanton event creates a $b$ anti-particle together with flux $2\pi$
out of the vacuum.
As noted above, this composite object has fermionic statistics. Thus, single fermion excitations
are created locally by instanton events: the surface theory has a statistical anomaly, and therefore, cannot be realized in a purely two-dimensional model.

Note that the above argument explains why $\sigma_{xy}$ must be an even integer for a purely two-dimensional bosonic insulator with no topological order, in accordance with the $K$-matrix
construction of Ref.~\onlinecite{Ashvin2d},
as well as general constraints on the consistency of Chern Simons terms on 3 manifolds with and without spin structures\cite{witten}.
Indeed, for a general integer level $k$ of the Chern-Simons theory (\ref{eq:CS2}),
$b$ bosons have
flux $-2 \pi/k$ bound to them and are transmutted to anyons with a statistical angle $\theta = -\pi/k$.\footnote{The 2d statistical angle should not be confused with the $\theta$-angle in 3d.}  An instanton tunnelling event involves the creation of flux $2 \pi$ (and the corresponding polarization charge $k$) 
together with  $k$ $b$-anti-particles.  This composite object has statistics $\theta = -\pi k$ and so $k$ must be an even integer in order to prevent single fermions from being created out of the vacuum. 

We also briefly note that the above argument is consistent with the fact that a two-dimensional fermionic insulator with no topological order can have any integer $\sigma_{xy}$. Indeed, suppose the charge carriers in the theory (\ref{eq:CS2})
are fermions $f$ with unit charge. These excitations again have a flux $-2 \pi/k$ dynamically bound to them and their statistics is transmutted to $\theta = \pi - \pi/k$. An instanton event involves the creation of 
flux $2\pi$ together with $k$ $f$-anti-particles. This composite object has statistics $\theta = \pi k^2 - \pi k $, which is bosonic as long as $k$ is an integer.  

Let us now return to the surface of the bosonic topological insulator with $\sigma_{xy} = 1$ and discuss how the anomaly in the 2d model
is resolved in three dimensions. Let's place a trivial vacuum in the region $z > 0$,  a bosonic topological insulator in the region $z < 0$ and an interface with $\sigma_{xy} = 1$ at $z = 0$.
We now couple the system to a fully three-dimensional compact $U(1)$ gauge field. The simplest thought experiment to carry out is to start with a neutral monopole in vacuum and let it pass through the surface of a bosonic topological insulator, picking up charge $-1$  and leaving a polarization charge $1$ on the surface. As discussed in the previous section, a neutral monopole in vacuum is a boson, likewise, a monopole with charge $-1$ inside the topological insulator
is a boson. Thus, statistics are conserved in the process considered. Alternatively, we can imagine starting with a neutral monopole in vacuum and letting it pass through the topological insulator surface while remaining
neutral. This process will leave a $b$-anti-particle on the surface together with a polarization charge $1$. Now, the neutral monopole in the bulk of the topological insulator is a fermion. However, the statistical interaction between the $b$-boson on the surface and the monopole in the bulk allows us to view this pair together as a boson. Thus, statistics are again conserved.

We can also look at the full 3d problem from the perspective of surface physics. It turns out that the electromagnetic response of the surface with $\sigma_{xy} = 1$ endows the excitations near the surface
with an effective 2d statistical interaction. Processes where two particles are exchanged in the plane of the surface over distances much larger than their separation from the surface receive a Berry's phase, which 
mimics two-dimensional statistics. Labelling the monopoles on the vacuum side of the interface as $m_+$ and monopoles on the topological insulator side of the interface as $m_-$, we obtain the following effective 2d exchange
statistics (see appendix \ref{app:sigmaxy}) 
\bea \theta(b,b) &=& - \frac{\pi \alpha^2}{1+ \alpha^2}, \quad \theta(m_+, m_+) =  \frac{\pi}{4 (1+ \alpha^2)}, \quad \theta(m_-, m_-) = \frac{\pi}{4 (1+ \alpha^2)} + \pi\label{eq:selfstat}\\
\theta(b, m_\pm) &=& \mp\frac{\pi}{1+\alpha^2}, \quad \theta(m_+, m_-) = -\frac{\pi}{2 (1+ \alpha^2)}\label{eq:mutualstat}\eea
Here, Eq.~(\ref{eq:selfstat}) summarizes the self-statistics and Eq.~(\ref{eq:mutualstat}) - the mutual statistics of excitations. We note that the $b$-particle preserves its self and mutual statistics as it passes through the interface,
so there is no need to distinguish $b$ excitations on the two sides of the interface. Note that all the statistical angles depend on the three-dimensional fine-structure constant $\alpha = \frac{e^2}{4 \pi}$  (we assume, for simplicity, that the dielectric constant and permeability of the topological insulator are the same as of vacuum). 
Note, in particular, that unlike in a theory where the electromagnetic field is purely two-dimensional, the effective statistics $\theta(b,b)$ of the $b$-particles near the surface is generally not fermionic; $\theta(b,b) \to - \pi$ only when $\alpha \to \infty$.  
 We also note that unlike in a purely two-dimensional theory, static $b$-particles will experience long-range $1/r$ Coulomb interactions mediated by the gapless bulk photon. Finally, the extra $\pi$ in the self-statistics of $m_-$ compared to $m_+$ reflects the intrinsic fermionic nature of monopoles in the topological insulator. 

Let us repeat the thought experiment above. Take a neutral monopole in vacuum ($m_+$) and let it tunnel through the interface turning into a neutral monopole $m_-$ and creating a $b$-anti-particle. Now, the 2d self-statistics of an anti-$b$ - $m_-$ composite is $\theta =  \theta(b,b) - \theta(b,m_-) + \theta(m_-,m_-) = \frac{\pi}{4(1+\alpha^2)}$, which turns out to be exactly the same as the self-statistics of $m_+$. Hence, statistics are again conserved! 

\subsection{Surface superfluid.}
\label{sec:sf}

Another surface phase proposed by VS is a superfluid where the global $U(1)$ symmetry is spontaneously broken but the time-reversal symmetry is preserved. The excitations of this phase include
a 2d gapless Goldstone mode and gapped superfluid vortices on the surface. 
An effective theory of this surface phase is most easily expressed in dual variables. The surface action $S = \int d^2 x d\tau L$ with
\beq L = \frac{1}{8 \pi^2 \rho_s} (\epsilon_{\mu \nu \lambda} \d_\nu a_{\lambda})^2 + i (a_{\mu} + \frac{A_{\mu}}{2}) j^+_{\mu}  + i (a_{\mu} - \frac{A_{\mu}}{2}) j^-_{\mu}
- \frac{i}{2\pi}   \epsilon_{\mu \nu \lambda} A_{\mu}\d_{\nu} a_\lambda  .\label{eq:Lsf}\eeq
Here, the superfluid current $J^{s}_{\mu}$ is expressed in terms of the dual gauge field $a_{\mu}$ as  $J^{s}_{\mu} = -\frac{1}{2 \pi} \epsilon_{\mu \nu \lambda} \d_{\nu} a_{\lambda}$. The dual gauge
field $a_{\mu}$ should not be confused with the electromagnetic gauge field $A_{\mu}$, which we treat at this stage as a non-compact external probe field.
Here $\rho_s$ denotes the superfluid
stiffness. The effective 2d theory involves two types of gapped superfluid vortices, $\psi_{\pm}$; $j^{\pm}$ are the corresponding vortex currents. The vortices are minimally coupled to $a_{\mu}$. Note that
under time-reversal, $\psi_\pm \to \psi^{\dagger}_\mp$. 


The only unusual property of the surface superfluid described by the action (\ref{eq:Lsf}), compared to an ordinary two-dimensional superfluid, is that the vortices $\psi_\pm$ formally carry a global $U(1)$ charge $\pm 1/2$. Thus, the 
total electromagnetic current,
\beq J^{EM}_{\mu} = -\frac{1}{2 \pi} \epsilon_{\mu \nu \lambda} \d_{\nu} a_{\lambda} + \frac12 (j^+_{\mu} - j^-_{\mu})  , \label{eq:JEMsf}\eeq
involves both the superfluid current and the vortex currents. Since the global $U(1)$ symmetry in a superfluid is spontaneousy broken, it is inappropriate to label its excitation by their global charge; we will discuss
a more physical distinction between the present surface superfluid and an ordinary purely two-dimensional superfluid shortly.

As  noted, we temporarily switch off the fluctuations of the external electromagnetic field $A_{\mu}$. Then the dual gauge field $a_{\mu}$ will be in the Coulomb phase, i.e. the dual photon will be gapless and will directly 
correspond to the superfluid Goldstone mode. The gapless $2+1$ dimensional photon mediates a logarithmic Coulomb interaction between static 2d surface vortices. We note that since flux $\phi$ of $a_{\mu}$
corresponds to a global charge $-\phi/2\pi$, pure flux-tunneling events of $a_{\mu}$ are prohibited by charge conservation. As a result, the dual gauge field does not go into a confined phase: the superfluid is stable. However,  events where 
flux $\phi = 2\pi$ is created together with a vortex $\psi^{\dagger}_+$ and an antivortex $\psi_-$ are allowed. Indeed, such events conserve both the global $U(1)$ charge and the vorticity of the superfluid.
We can write the corresponding term in the Lagrangian as,
\beq \Delta L =  \lambda m^{\dagger} \psi^{\dagger}_+ \psi_- + h.c.  ,\label{eq:Ltun}\eeq
where the monopole operator $m^{\dagger}$ creates flux $2\pi$ of $a_{\mu}$ and $\lambda$ is a coupling constant. It is well-known that $m^{\dagger}$ acquires a finite expectation value in the 2d Coulomb phase.
In fact, since $m^{\dagger}$ carries charge $-1$ under the global $U(1)$ symmetry, it serves as the superfluid order parameter and may be identified with the physical boson operator $b$. Moreover, we may schematically replace $m^{\dagger}$ in Eq.~(\ref{eq:Ltun}) by its expectation value,
\beq \Delta L \to \lambda \langle m^{\dagger} \rangle \psi^{\dagger}_+ \psi_- + h.c.    .\label{e:Ltun2}\eeq
Hence, the term (\ref{eq:Ltun}) induces tunneling between the gapped $\psi_+$ and $\psi_-$ vortices; energy eigenstates will be superpositions of $\psi_+$ and $\psi_-$. Note that time-reversal maps the vortex
$\psi^{\dagger}_+$ to an antivortex $\psi_-$. Hence, in the absence of an extra particle-hole symmetry, there is no reason for $\psi_+$ and $\psi_-$ vortices to have the same energy and so a finite tunneling
between them generally will not give rise to an equal weight superposition. 

Let us now switch on the fluctuations of $A_{\mu}$, first treating it as a non-compact two-dimensional gauge-field. The effective 2d theory (\ref{eq:Lsf}) then becomes fully gapped - indeed, when the global $U(1)$ symmetry of a superfluid is gauged, the Goldstone mode
disappears and the external electromagnetic field becomes Higgsed. The effect of the mutual Chern-Simons term $\frac{i}{2\pi}   \epsilon_{\mu \nu \lambda} A_{\mu}\d_{\nu} a_\lambda$ in Eq.~(\ref{eq:Lsf}) is to attach flux
$2 \pi$ of $A$ to $\psi^{\dagger}_\pm$: the superfluid vortices become magnetic flux tubes. Moreover, a flux $\pm \pi$ of $a$ is attached to $\psi^{\dagger}_\pm$; Eq.~(\ref{eq:JEMsf}) indicates that this flux
corresponds to an electric polarization charge $\mp 1/2$ under $A$. Thus, the $\psi_\pm$ flux tubes, which couple minimally to $A_\mu$ with charges $Q=\pm1/2$, as in Eq.~(\ref{eq:Lsf}), become overall electrically neutral, in accordance with the fact that electric charge is Debye screened in a superconductor. 

As in an ordinary 2d superconductor, the interaction between two static flux tubes is exponentially screened - the flux tubes are local excitations. However, the flux attachment discussed above endows the flux-tubes with a statistical interaction: both $\psi_+$ and 
$\psi_-$ acquire fermionic self-statistics. The mutual statistics between $\psi_+$ and $\psi_-$ are bosonic. Indeed, we can think of the action (\ref{eq:Lsf}) as a Chern-Simons theory for the two-component gauge field $(a_{\mu}, A_{\mu})$ with the $K$-matrix, $K = \left(\begin{array}{cc} 0 & 1\\1& 0\end{array}\right)$. The $\psi_{\pm}$ vortices carry charges $(1, \pm 1/2)$ under this two-component gauge field; the 
aforementioned statistics immediately follow.

Note that the tunelling operator (\ref{eq:Ltun}) converts a $\psi_-$ flux-tube to a $\psi_+$ flux-tube: the statistics
 is preserved in the process. Hence, the statistics of the energy eigenstates, which are superpositions of $\psi_+$ and $\psi_-$ will likewise be fermionic.

We conclude that the key signature of the superfluid on the surface of a topological insulator is that upon coupling to a 2d electromagnetic field, its flux-tubes possess fermionic statistics. This is in contrast to an
ordinary time-reversal invariant purely two-dimensional superfluid, whose flux tubes are bosons. The fermionic statistics of flux-tubes directly imply that a putative purely 2d system with the same properties suffers from a statistical anomaly and so cannot exist. Indeed, if the effective theory (\ref{eq:Lsf}) emerges from a purely 2d lattice model, we should be able to promote the electromagnetic field $A_{\mu}$ to a compact gauge field. An instanton of $A_{\mu}$ creates a flux $2\pi$ excitation out of the vacuum. The only such excitations in the theory are the fermionic flux-tubes. 
Thus, single fermions are created out of the vacuum during
instanton events and a purely two-dimensional model is inconsistent.

How is the statistical anomaly resolved in the full 3d theory in the present case? The electromagnetic field in the bulk of the system is in the Coulomb phase. Nevertheless, the magnetic field must still penetrate the
surface in the form of tubes with quantized flux. The ``demagnetization effects" in the bulk lead to a $V(r) = \frac{1}{\alpha r}$ interaction between static flux-tubes. Moreover, the magnetic field profile of a single flux-tube on the surface now has an algebraic $1/r^3$ tail, instead of falling off exponentially, as in a purely 2d system.
Likewise, the electric charge cloud that screens the $\pm 1/2$ charge of a $\psi_\pm$ vortex has a density profile with a $1/r^3$ tail. It turns out that these tails fall off sufficiently quickly for the statistics of flux-tubes to 
remain well-defined and fermionic. 

Thus, as a monopole passes from the vacuum outside the topological insulator through the interface, it changes its statistics from bosonic to fermionic and excites a fermionic flux-tube on the surface. So, the statistics are conserved!

Before we conclude this section, we would like to stress that the fermionic statistics acquired by flux tubes upon gauging the $U(1)$ symmetry are not directly related to the statistics of global superfluid vortices in
the absence of an external gauge field.  VS have claimed that it is also appropriate to think of the statistics of global superfluid vortices on the surface of a topological insulator as fermionic. As we discuss in appendix \ref{app:vort}, this claim is correct only when the system has an additional particle-hole symmetry. On the other hand, the fermionic statistics of flux-tubes are completely robust to particle-hole symmetry breaking. Moreover, as emphasized in appendix \ref{app:vort}, the two effects have different dynamical origins.

\subsection{Surface phase with $Z_2$ topological order.}

Yet another surface phase proposed by VS is fully gapped, respects the $U(1)\ltimes Z^T_2$ symmetry and carries an intrinsic  2d $Z_2$ (toric code) topological order. The excitations of this phase are $e$ and 
$m$ anyons carrying electric and magnetic charge respectively under the emergent $Z_2$ gauge field. $e$ and $m$ have bosonic self-statistics and are mutual semions. The bound state of $e$ and $m$
is a fermion. The $e$ and $m$ anyons transform trivially under time-reversal, but both carry charge $1/2$ under the global $U(1)$ symmetry. This property is somewhat unusual; standard two-dimensional bosonic states with $Z_2$ topological order,
global $U(1)$ and time-reversal symmetries, have just one (or none) of the anyons carrying charge $1/2$. We will now argue that precisely this property makes the present phase impossible to realize in a purely 
two-dimensional system.

Again, if the state with the above properties can be realized in a purely 2d lattice model, we should be able to couple the model to a compact 2d $U(1)$ gauge field. Let us imagine a flux-tunneling event in such a coupled model. We start with an initial state with no magnetic flux and no charge, nucleate flux $2\pi$ through 
a single plaquette at the origin and let it expand to a smooth distribution. Since the system is time-reversal invariant, $\sigma_{xy} = 0$, and so the flux carries no electric charge. Thus, any local quasi-particle created during 
the instanton tunneling event must be neutral. Now, both $e$ and $m$ charge $1/2$ anyons pick up a Berry's phase $\pi$ upon encircling the flux in the final state, but acquire no Berry's phase upon encircling the origin in the initial state. In a local model an instanton tunneling event cannot change the Berry's phase acquired by a distant quasiparticle upon encircling the spatial location of the event. The only possible resolution of the above puzzle is that the instanton event must create a quasiparticle, which has mutual semionic 
statistics with both $e$ and $m$ anyons, compensating the $\pi$ Berry's phase due to the magnetic flux. This quasiparticle must be identified with the neutral fermion $f = e m^{\dagger}$. Hence, the instanton event creates a neutral $f$ fermion together with flux $2 \pi$. Since $f$ is neutral, its statistics are not affected by the $2 \pi$ flux and remain fermionic. Thus, single fermions are created out of the vacuum during such instanton tunneling events.  This is not possible in a local 2d model, which is thus seen to possess a statistical anomaly.


We note in passing that, as pointed out by VS, a toric code with both $e$ and $m$ anyons carrying charge $1/2$ can be realized strictly in 2d, provided that the system has a Hall-conductivity $\sigma_{xy} = 1$.  This 2d system breaks time reversal symmetry and is described by a $K$ matrix construction, with $K = \left(\begin{array}{cc} 0 & 2\\2& 0\end{array}\right)$ and the charge vector $t = (1, 1)$. The relationship between this 2d state and the $\sigma_{xy}=0$ surface state can be understood by considering a slab of topological insulator of a large but finite thickness, with the $\sigma_{xy} = 0$ toric code state on the top surface, and the non topological $\sigma_{xy}=1$ state on the bottom surface. The entire slab viewed as a 2d system realizes the toric code with $\sigma_{xy} =1$. It follows that an interface between the toric code state and the $\sigma_{xy}=1$ state on the surface of the topological insulator will exhibit edge states identical to those of the 2d $\sigma_{xy}=1$ state.

Since a $K$-matrix construction exists, the 2d theory should be free of anomalies. Let us explicitely check this. As usual, let's couple the 2d model to a compact $U(1)$ gauge field and imagine an instanton event which creates flux $2\pi$ out of the vacuum. Since $\sigma_{xy} = 1$, this flux carries a polarization charge $1$. As for the toric code on the surface of the topological insulator, the instanton also creates an $e m$ bound state, whose intrinsic semionic mutual statistics with the charge $1/2$ $e$ and $m$ anyons cancels the $\pi$ Berry's phase acquired by the latter upon encricling the $2 \pi$ flux. However, in the present case, the $e m$ bound state carries charge $-1$ in order to conserve the total charge. Thus, the instanton creates an $e m$ bound state with charge $-1$ together with flux $2 \pi$. The attached flux transmutes the intrinsic fermionic statistics of $e m$ to bosonic. Thus, the statistics are conserved in the process and the theory is anomaly free.



Returning to the bosonic topological insulator with the time-reversal invariant toric code surface, in the present case, the anomaly is resolved very directly in three dimensions: the monopole simply passes through the surface of the topological insulator changing its statistics and leaving an $f$ fermion on the surface.
Thus, the fermion parity is conserved. Note that both the monopole and the $f$ particle are electrically neutral, so there is no long-range statistical interaction between them mediated by the 3d photon.

\subsection{General constraints on the surface.}
The discussion in the previous section immediately generalizes to a proof of the statement that if the surface of a bosonic topological insulator is neither gapless, nor spontaneously breaks the symmetry it must
possess topological order. Indeed, suppose the surface is fully gapped and does not break the time-reversal symmetry. Then the surface Hall-conductivity $\sigma_{xy} = 0$. Now, let's tunnel a neutral monopole
across the interface between the vacuum and the topological insulator. Since $\sigma_{xy} = 0$, no polarization charge is induced on the surface. This means that any excitation created on the surface during the tunneling
process is neutral. Thus, this excitation will possess no statistical interaction with the monopole in the bulk. Now, the neutral monopole in the bulk is a fermion. Thus, the quasiparticle left on the 
surface is likewise a fermion. Hence, the gapped surface state must support fermionic excitations. Since we are dealing with a system made out of bosons, the presence of fermionic quasiparticles implies that the surface phase has intrinsic 2d topological order. QED.

The above argument not only proves that the gapped, symmetry respecting surface of the topological insulator must support intrinsic topological order, but also places some constraints on the order allowed.
In general, it may be convenient to label SPT phases by their gapped, symmetry respecting topologically ordered surface states. Such a label will carry the information about the intrinsic topological order
(fusion rules, braiding statistics), as well as the quantum numbers of anyon excitations under the global symmetry. Of course, many different topologically ordered surface states may be realized; in fact, one can always
``paint" an additional layer of a purely two-dimensional topologically ordered phase on top of the surface. A possible way to arrange the different topologically ordered states is by their quantum dimension ${\cal D}$. 
Thus, we may use the surface state(s) with the lowest quantum dimension to label an SPT phase. (There may still be several such states allowed). 

We note that {\it any} surface state of an SPT that has intrinsic topological order, apart from the realization of global symmetry, has to be identical to one of the allowed strictly 2d topological orders.  This follows immediately from the slab construction mentioned above, in which the top surface is the state in question and the bottom surface is in a gapped global symmetry violating topologically trivial state.  Since both the bulk of the SPT phase and the bottom surface support no topologically non-trivial excitations, the topological order of the slab as a whole is just given by the topological order on the top surface. This proves that the braiding statistics on the surface of an SPT phase must be realizable strictly in 2d. Note that since the bottom surface in the above construction breaks the symmetry, this argument makes no statement about the realization of symmetry in the topologically ordered state on the top surface.

Let us now identify the symmetry respecting topologically ordered surface state(s) of the bosonic topological insulator with the lowest quantum dimension. One allowed topologically ordered surface state is the toric code with charge $1/2$ anyons, discussed in the previous section. This state
has a quantum dimension ${\cal D} = 2$. It is known that there are no time-reversal invariant topological orders with a smaller quantum dimension.\cite{ZWang} Moreover, the only other time-reversal invariant topological
order with ${\cal D} = 2$ is the double-semion theory.\cite{ZWang} However, as argued above, the topologically ordered state on the surface must posses fermionic quasiparticles. The double-semion theory supports no fermionic excitations. Thus, we conclude that the toric code with charge $1/2$ anyons is the unique symmetry respecting surface state of the bosonic topological insulator with the smallest quantum dimension.


\section{Relation to the cohomology classification}
\label{sec:coho}
In this section, we briefly comment on the potential place of the bosonic topological insulator phase considered in the present paper within the cohomology classification of Ref.~\onlinecite{WenCohoBoson}. A similar discussion has been given by VS. 

The cohomology classification predicts a $Z^2_2$ structure for 3d phases with the bulk symmetry group $U(1) \ltimes Z^T_2$. This means that there 
are three non-trivial phases, $g_1$, $g_2$, $g_3$ where the third one, ``$g_3 = g_1 + g_2$," can be thought of as a weakly interacting ``mixture" of the first two. Moreover, the cohomology technology
predicts a $Z_2$ classification for phases with just the $Z^T_2$ symmetry and no nontrivial phases with just the $U(1)$ symmetry. This implies that one of the three non-trivial phases with 
$U(1) \ltimes Z^T_2$ symmetry, say $g_1$, is just the non-trivial $Z^T_2$ phase, i.e it does not involve the $U(1)$ symmetry in any interesting way. In particular, if we start with $g_1$ and explicitely break the 
$U(1)$ symmetry, the resulting state is still protected by $Z^T_2$ and cannot be smoothly connected to a product state. On the other hand, since $g_3 = g_1 + g_2$,  out of the remaining two phases,
$g_2$ and $g_3$, one (say $g_2$) must become trivial under the $Z^T_2$ classification if $U(1)$ is explicitely broken, while the other ($g_3$) remains non-trivial. The phase that we refer to as the bosonic topological
insulator in this paper is $g_2$: it is unstable whenever either the $U(1)$ or the $Z^T_2$ symmetry is broken in the bulk. VS have also constructed an effective theory for the pure $Z^T_2$ phase $g_1$, and thereby, for the phase $g_3$.   However, we do not consider these phases in the present work.

We also note that VS have proposed an additional phase with just the $Z^T_2$ symmetry, which falls outside the cohomology classification of Ref.~\onlinecite{WenCohoBoson}. The time-reversal symmetry breaking surface state of this phase has a thermal Hall response with $\kappa_{xy} = 4$, so that a domain wall
on the surface between two time-reversal conjugate regions with $\kappa_{xy} = \pm 4$ supports 8 chiral gapless modes ($c_L - c_R = 8$).  This gapless domain wall is identical to the edge state of a 2d $\mathbb{E}_{8}$ integer quantum Hall state of bosons with $\kappa_{xy} = 8$. Recently, several explicit constructions of this 3d phase 
have appeared\cite{SenthilLayered, AshvinExact} (including an exactly solvable model in Ref.~\onlinecite{AshvinExact}). Again, we do not consider this phase (and its mixtures with ``conventional" cohomology phases) in this paper.

\section{Conclusion}
In the present paper we have identified the statistical Witten effect as the bulk signature of three-dimensional bosonic topological insulators, protected by the symmetry group $U(1)\ltimes Z^T_2$. We have shown that this effect immediately implies that the surface physics of the topological insulator cannot be fully realized in a local two-dimensional model. The statistical Witten effect also implies that if the surface is neither gapless, nor spontaneously breaks the symmetry, it must be topologically ordered. Moreover,
we have demonstrated that the surface phases of the bosonic topological insulator inferred by Vishwanath and Senthil\cite{VS} are consistent with the statistical Witten effect in the bulk.

A reader may ask, how do the authors know that time reversal respecting insulators exhibiting the statistical Witten effect can exist?  In a companion paper,\cite{Forthcoming} we will provide an explicit field-theoretic, lattice-regularized, construction
of a time-reversal invariant bosonic insulator, which displays the statistical Witten effect in its bulk and realizes the surface phases of VS. The first step of our construction involves a parton decomposition of the microscopic bosons, which yields a phase with an emergent gapless $u(1)$ gauge field and deconfined parton and monopole excitations. (The emergent gauge field and its monopoles should not be confused with the external electromagnetic gauge field). In the second step, we form certain dyon bound states of monopoles and partons, and by dyon condensation   
drive a confinement transition to a fully gapped phase with no intrinsic topological order. This phase has all the bulk and surface properties of a bosonic topological insulator discussed in the present paper.

In this work we have pursued the line of attack,\cite{LevinGu,Wenduality} where symmetry protected topological phases are differentiated by ``weakly gauging" their global symmetry. As the present paper was being completed,   significant broad progress on the same front was reported in Ref.~\onlinecite{WenGauge}, which utilized this general approach to distinguish SPT phases within the cohomology classification\cite{WenCohoBoson} with various global symmetries. However, Ref.~\onlinecite{WenGauge} did not consider any phases with time-reversal symmetry, and so the $U(1) \ltimes Z^T_2$ topological insulator phase, to which the present paper is devoted, fell outside its scope. In fact, we do not know how to ``gauge" the time-reversal symmetry. Thus, we do not have a completely general tool to diagnose SPT phases whose global symmetry group contains time-reversal. Fortunately, for the topological insulator phase considered in this paper, it was sufficient to gauge just the $U(1)$ part of the symmetry to expose the statistical Witten effect; time-reversal remained as a global symmetry of the resulting weakly fluctuating $U(1)$ gauge theory. However, there are phases within the cohomology classification, which appear to involve the time-reversal symmetry in a more ``active" manner. For instance, the classification predicts a single non-trivial SPT phase in three dimensions with just the time-reversal symmetry. Although a surface theory for this phase was proposed by VS, a bulk signature is currently lacking. Yet a bulk diagnostic would be greatly desirable, since the surface termination is generally not unique.   
Moreover, it is often not immediately obvious why a given surface phase cannot be realized in a purely two-dimensional system. As we have seen with the example of the ordinary/statistical Witten effect in fermionic/bosonic topological insulators, a bulk signature may be the key to identifying (and resolving) the surface ``anomaly."


\acknowledgements
We would like to thank A.~Vishwanath, T.~Senthil, X.~G.~Wen, C.~Xu , T.~Grover, G.~Cho, C.~Nayak and E.~Witten for stimulating discussions. Thanks also to the Aspen Center for Physics where some of this work was initiated.  This research was supported in part by the National Science Foundation under Grant No. NSF PHY11-25915,  
DMR-1101912 (M.P.A.F.) and DMR 0906175 (C.L.K.), by the Caltech Institute of Quantum Information and Matter, an NSF Physics Frontiers Center with support of the Gordon and Betty Moore Foundation (M.P.A.F.) and by a Simons Investigator award from the Simons Foundation (C.L.K.). 

\appendix

\section{Statistical interactions near the $\sigma_{xy} = 1$ surface of a bosonic topological insulator.}
\label{app:sigmaxy}
In this appendix, we study the surface of the bosonic topological insulator with $\sigma_{xy} = 1$ when the system is coupled to a weakly fluctuating 3d compact electromagnetic gauge field
$A_{\mu}$. We show that bosons $b$ and monopoles $m$ moving in the plane of the surface experience an interaction, which mimics two-dimensional fractional exchange 
statistics. 

We  begin with the following action,
\beq S = \frac{1}{4 e^2} \int d^4x (F_{\mu \nu} - 2 \pi M_{\mu \nu})^2 + \frac{i}{4 \pi} \int_{z = 0}  d^3 x_\parallel \epsilon_{i j k} A_{i} \d_{j} A_{k}
+ i \int d^4 x A_{\mu} J^{b}_\mu + S^0_B \label{eq:Ssurf3d}\eeq
Here, the region $z >0$ is occupied by the trivial vacuum, the region $z < 0$ by the topological insulator and the interface lies in the  $z  = 0$ plane. The Greek indices run over $\tau, x, y, z$, while the Latin indices run over $\tau, x, y$; $x_\parallel = (\tau, x ,y)$ labels the coordinates in the plane of the interface. The Chern-Simons term for $A$ encodes the $\sigma_{xy} = 1$
electromagnetic response of the surface.  $J^b_{\mu}$ is the 3+1 dimensional 
boson current. $M_{\mu \nu}$ represents the Dirac strings of magnetic monopoles, so that the 3+1 dimensional monopole current is given by  
$J^m_{\mu} = - \frac{1}{2} \epsilon_{\mu \nu \lambda \sigma} \d_{\nu} M_{\lambda \sigma}$. $S^0_B$ is a Berry's phase term, which endows the monopoles in the topological insulator with fermionic statistics. We will
give an explicit expression for this term below.

Note that we have not ``compactified" the Chern-Simons term; here, for simplicity, we will not expicitely consider events where
monopoles tunnel through the surface. We can then separate monopoles into those in the trivial vacuum ($m_+$) and in the topological insulator ($m_-$), and write $M_{\mu \nu} = M^+_{\mu \nu} + M^-_{\mu \nu}$, 
with the corresponding monopole currents $J^{m \pm}_{\mu} =  - \frac{1}{2} \epsilon_{\mu \nu \lambda \sigma} \d_{\nu} M^\pm_{\lambda \sigma}$. We choose a gauge where the Dirac strings of $m^+$ and $m^-$ 
do not pass through the interface. More specifically, we take the Dirac string of a static $m^+$ monopole to run along the $+z$ direction and the Dirac string of a static $m^-$ monopole to run along the $-z$ direction:
\bea M^{\pm}_{i j} &=& - \epsilon_{i j k} \int d^4 x' S^{\pm}(x -x') J^{m \pm}_{k}(x'), \quad M^{\pm}_{i z} = 0 \label{eq:strings}\\
S^{\pm}&=& \pm \theta(\pm z) \delta(x) \delta(y) \delta (\tau)\eea
Then, $M^{+}, J^{m +}$ ($M^-$, $J^{m -}$) vanish as long as $z < 0$ ($z > 0$). 

It is now simple to give an expression for $S^0_B$:
\beq S^0_B =  \pi i \int d^4 x d^4 x' J^{m -}_{i} (x) \epsilon_{i j k}  \d_{j}  D_3(x_\parallel - x'_\parallel) J^{m -}_{k}(x')  \label{eq:SB0}\eeq
Here, $D_3(x_\parallel) = (4 \pi |x_\parallel|)^{-1}$, is the $2+1$ dimensional propagator. Eq.~(\ref{eq:SB0}) can also be rewritten as
\beq S^0_B = \pi i \int d^3 x_{\parallel} d^3 x'_{\parallel} J^{m - \parallel}_{i} (x_{\parallel}) \epsilon_{i j k}  \d_{j}  D_3(x_\parallel - x'_\parallel) J^{m - \parallel}_{k}(x'_\parallel)  \label{eq:SB02d}\eeq
where 
\beq J^{m \pm \parallel}_{i} (x_\parallel) = \int dz J^{m \pm}_{i} (x_\parallel, z) \label{eq:Jpar}\eeq
is the projection of the $3+1$ dimensional monopole current onto the $z = 0$ plane. Eq.~(\ref{eq:SB02d})  gives the familiar Berry's phase
for a 2d fermion with current $J^{m - \parallel}_i$. Now, if we consider a process where two 3d monopoles $m^-$ are exchanged, their projections onto the $z = 0$ plane are also exchanged. Hence, Eq.~(\ref{eq:SB02d}) accurately captures the full $3+1$ dimensional ``intrinsic" fermionic exchange statistics of $m^-$ monopoles. (Here, we do not consider configurations where two $m^-$ monopoles simultaneously have the same $x,y$ coordinates and so have the same projection onto the $z = 0$ plane. Such configurations form a set of measure zero.) 

We next integrate out the gauge field $A_{\mu}$ in the action (\ref{eq:Ssurf3d}) to determine the full statistical interaction between the $b$ and $m^\pm$ particles. To do so, it is convenient to introduce an auxilliary $2+1$ dimensional gauge field $\alpha_{i}$ living on the interface and rewrite (\ref{eq:Ssurf3d}) as,
\bea S &=& \frac{1}{4 e^2} \int d^4x (F_{\mu \nu} - 2 \pi M_{\mu \nu})^2  + i \int d^4 x A_{\mu} J^{b}_\mu + S^0_B \nn\\
&+& \int  d^3 x_\parallel \left(- \frac{i}{4 \pi} \epsilon_{i j k } \alpha_{i} \d_{j} \alpha_{k} + \frac{i}{2 \pi} A_{i} \epsilon_{i j k} \d_{j} \alpha_{k}\right)
\label{eq:Ssurfalpha}\eea
The integral over $A_{\mu}$ is now easy to perform, giving
\beq S_{\mathrm{eff}} = S^{3+1}_{\mathrm{eff}} + S^0_B + S^\alpha \label{eq:Seff}\eeq
where
\bea S^{3+1}_\mathrm{eff} &=& \frac{1}{2} \int d^4x d^4 x' \left( J^b_{\mu}(x) D_4(x-x') J^b_{\mu}(x') + \frac{(2 \pi)^2}{e^2} J^m_{\mu}(x) D_4(x-x') J^m_{\mu}(x')\right) \nn\\
&+& 2 \pi i \int d^4x d^4x' J^b_{\mu}(x) D_4(x-x') \d_{\nu} M_{\mu \nu}(x') \label{eq:S3p1}\eea
and
\bea S^\alpha &=& -\frac{i}{4 \pi} \int d^3 x_\parallel \, \alpha_{i} \epsilon_{i j k} \d_{j} \alpha_{k} + \frac{e^2}{2 (2 \pi)^2} \int d^3 x_{\parallel} d^3 x'_{\parallel} \alpha_{i}(x_\parallel)(-\d^2_\parallel 
\delta_{i j} + \d_{i} \d_{j}) D_4(x_\parallel - x'_\parallel) \alpha_{j}(x'_\parallel) \nn\\
&+& i \int d^3 x_{\parallel} ' \alpha_{i}(x_\parallel) j^{\alpha}_{i}(x_\parallel) 
\label{eq:Salpha}\eea
with
\beq j^{\alpha}_{i}(x_\parallel) = \int d^4 x' \epsilon_{i j k} \d_{j} D_4(x_\parallel - x') \left(\d_l M_{k l}(x') - \frac{i e^2}{2 \pi} J^b_k (x')\right)\eeq
Here, $D_4(x) = (4 \pi^2 |x|^2)^{-1}$, is the $3+1$ dimensional propagator. The effective action $S^{3+1}_{\mathrm{eff}}$ represents the standard bulk interaction between charges and monopoles mediated by the gapless photon, while $S^\alpha$ captures the 
modification of the interaction due to the interface. 

Let us first discuss $S^{3+1}_{\mathrm{eff}}$. The first two (real) terms
in Eq.~(\ref{eq:S3p1}) give rise to the $1/r$ Coulomb interaction between static charges/monopoles. The last (imaginary) term in Eq.~(\ref{eq:S3p1}), which we denote as $S^{3+1}_B$, encodes the bulk statistical interaction between charges and monopoles. Explicitely, 
\beq S^{3+1}_{B} =  - 2 \pi i \int d^4 x d^4 x' J^b_i(x) \epsilon_{i j k} \left(\d_j K^+(x-x') J^{m +}_k(x') + \d_j K^-(x-x') J^{m -}_k(x')\right) \label{eq:S3p1B}\eeq
where
\beq K^{\pm}(x_\parallel, z) = \pm \frac{1}{4 \pi^2 |x_\parallel|}\left(\frac{\pi}{2} \pm \tan^{-1}\left(\frac{z}{|x_\parallel|}\right)\right) \label{eq:Kpm}\eeq
We note that the seeming difference between $K^+$ and $K^-$ corresponds to a gauge choice; a replacement of $K^-$ by $K^+$ in Eq.~(\ref{eq:S3p1B}) modifies the action by a multiple of $2 \pi i$.
Here, we are interested in configurations where the separation between the excitations along the surface, $|x_\parallel|$, is much larger than their distance to the surface $z$. In this ``2d" limit, we may
set $z$ in Eq.~(\ref{eq:Kpm}) to zero. (A finite value of $z$ gives corrections to the ``2d" statistical interaction, which are less relevant in the RG sense.) We then obtain,
\beq S^{3+1}_B = -\pi i \int d^3 x_\parallel d^3 x'_\parallel J^{b \parallel}_i (x_\parallel) \epsilon_{i j k} \d_j D_3(x_\parallel - x'_\parallel) (J^{m +\parallel}_k(x'_\parallel) - J^{m - \parallel}_k(x'_\parallel)) \label{eq:S3p1sem}\eeq
with $J^{b \parallel}_i$ defined analogously to Eq.~(\ref{eq:Jpar}). Eq.~(\ref{eq:S3p1sem}) implies that in the absence of a $\sigma_{xy} = 1$ surface, charges and monopoles behave as mutual semions, when
moving in a 2d plane. This is the expected result.

We now turn our attention to the modification of the bulk statistics by the interface, described by the $S^\alpha$ term (\ref{eq:Salpha}). Integrating the gauge field $\alpha_i$ out, we obtain,
\beq S^{\alpha}_{\mathrm{eff}} = \frac{1}{2} \int d^3 x_\parallel  d^3 x'_\parallel j^{\alpha}_{i}(x_\parallel) D^{\alpha}_{ij}(x_\parallel - x'_\parallel) j^{\alpha}_{j}(x'_\parallel) \label{eq:Salphaeff} \eeq
with $D^\alpha$ - the propagator of the $\alpha$ field, explicitely given by
\beq D^\alpha_{i j}(x_\parallel) = \frac{2 \pi}{1 + \alpha^2}(-i \epsilon_{i j k} \d_k D_3(x_\parallel) + 2 \alpha \delta_{i j} D_4(x_\parallel))\eeq
Here, we work in the gauge $\d_i \alpha_i = 0$. After some algebra, we find that  in the ``2d" limit described above, the imaginary (Berry's phase) part of Eq.~(\ref{eq:Salphaeff}) simplifies to,
\beq S^{\alpha}_{B} = i \int d^3 x_\parallel d^3 x_\parallel' \vec{J}^T_i(x_\parallel) \Theta^{\alpha} \epsilon_{i j k} \d_j D_3(x_\parallel - x'_\parallel) \vec{J}_k(x'_\parallel) \label{eq:SBalpha}\eeq
with $\vec{J} = (J^{b \parallel}, J^{m + \parallel}, J^{m - \parallel})$ and
\beq \Theta^{\alpha} = \frac{\pi}{1+\alpha^2}\left(\begin{array}{ccc} -\alpha^2 & \alpha^2/2 & -\alpha^2/2 \\ \alpha^2/2 & 1/4 & -1/4 \\ -\alpha^2/2 & -1/4 & 1/4\end{array}\right)\eeq
After combining Eq.~(\ref{eq:SBalpha}) with the  bulk Berry's phases, Eqs.~(\ref{eq:SB02d}), (\ref{eq:S3p1sem}), we find that the total Berry's phase for 2d exchange processes is given by,
\beq S_B = i \int d^3 x_\parallel d^3 x_\parallel' \vec{J}^T_i(x_\parallel) \Theta \epsilon_{i j k} \d_j D_3(x_\parallel - x'_\parallel) \vec{J}_k(x'_\parallel) \label{eq:SBtot}\eeq
with
\beq \Theta = \frac{\pi}{1+\alpha^2}\left(\begin{array}{ccc} -\alpha^2 & \,\,-1/2 & \,\,1/2\\ -1/2 & \,\,1/4 & \,\,-1/4 \\ 1/2 & \,\,-1/4 & \,\,1/4 + (1 + \alpha^2)\end{array}\right) \label{eq:Thetafin}\eeq
The effective statistical angles in Eqs.~(\ref{eq:selfstat}), (\ref{eq:mutualstat}) immediately follow. We can now imagine a process discussed in the main text, where an $m^+$ monopole tunnels through the interface, turning
into an $m^-$ monopole and a $b$ anti-particle, and leaving a polarization charge $1$ on the surface. Such a process can be thought off as a creation of an excitation with quantum numbers $(-1, -1, 1)$, corresponding to 
entries in the vector $\vec{J}$. One can use Eq.~(\ref{eq:Thetafin}) to check that the $(-1,-1,1)$ excitation has bosonic self-statistics and trivial mutual statistics with all the other excitations. Therefore, monopole tunneling events preserve the statistics and the theory is consistent.

\section{Statistics of superfluid vortices on the surface of a bosonic topological insulator.}
\label{app:vort}
This appendix is devoted to the statistics of vortices on the superfluid surface of a 3d bosonic topological insulator in the absence of a fluctuating electromagnetic gauge field. As discussed in section \ref{sec:sf}, 
the superfluid surface is described by the effective theory in Eqs.~(\ref{eq:Lsf}),(\ref{eq:Ltun}). Here, we turn off the external electromagnetic field $A_{\mu}$. 

As superfluid vortices have a long-range logarithmic interaction, the notion of statistics here is formal. For the present purposes, we define statistics as the Berry's phase in the imaginary time path integral accumulated during an adiabatic exchange. With this formal definition, as long as the tunneling between the $\psi_+$ and $\psi_-$ vortices in Eq.~(\ref{eq:Ltun}) is switched off, the statistics of $\psi_{\pm}$ are clearly bosonic. Any ``statistical interaction" between superfluid vortices must, therefore, come from this tunneling term (\ref{eq:Ltun}). In contrast, once fluctuations of the external electromagnetic field are switched on, vortices acquire fermionic statistics even when the tunneling term (\ref{eq:Ltun}) is absent. Moreover, as already noted, once the theory is gauged, a weak tunneling $\lambda$ between the $\psi_+$ and $\psi_-$ flux-tubes does not affect their statistics. Hence, the fermionic statistics of flux-tubes and any potential statistics of superfluid vortices have different dynamical origins.

Let us now turn on a weak tunneling term (\ref{eq:Ltun}) and carefully examine the statistics of resulting superfluid vortex excitations. We can think of an isolated vortex as a two-level system ($\psi_+$, $\psi_-$) with an energy splitting $2 \Delta$ related to the difference of bare energies of $\psi_+$ and $\psi_-$ vortices. As we already pointed out, in the absence of an additional particle-hole symmetry, $\Delta$ will generally be non-vanishing. Our two-level system also has a tunneling amplitude $\lambda \langle m^{\dagger} \rangle$ related to the local expectation value of the monopole operator $m$. Hence, we can think of the vortex as a spin in a magnetic field $\vec{b} = (\lambda Re\langle m^{\dagger}\rangle, \lambda Im \langle m^{\dagger} \rangle, \Delta)$. 

As we already noted, $m^{\dagger}$ may be interpreted as the local order parameter of the superfluid. According to this interpretation, the phase of $\langle m^{\dagger} \rangle$ should wind by $2 \pi$ around a superfluid vortex. This can be confirmed by an explicit calculation of the monopole expectation value in our dual gauge theory description.\cite{MS} 

Now, imagine starting with vortex 1 at $\vec{x} = (a,0)$ and vortex  2 at $\vec{x} = (-a,0)$, and performing a counter-clockwise exchange operation.  Each vortex ``spin," $\xi_{1,2}$, will see the effective ``magnetic field," $\vec{b}_{1,2}$, generated by the phase of the other vortex. We may write, $\vec{b}_1(\tau) = (\lambda |\langle m \rangle| \cos \theta(\tau), \lambda |\langle m \rangle| \sin \theta (\tau), \Delta)$ and $\vec{b}_2(\tau) = (-\lambda |\langle m \rangle| \cos \theta(\tau), -\lambda |\langle m \rangle| \sin \theta (\tau), \Delta)$, with $\theta(\tau)$ evolving with time from $0$ to $\pi$, so that $\vec{b}_1$ and $\vec{b}_2$ exchange during the process. If we start with both vortex spins in their respective ground states, the system picks up a Berry's phase equal to the area on the unit sphere traced out by $\vec{b}_1$ and $\vec{b}_2$, namely $S_B = i \pi (1 - \Delta/\sqrt{\lambda^2|\langle m \rangle|^2 + \Delta^2})$. On the other hand, if both vortex spins are in their excited states, an opposite Berry's phase is picked up. Hence, as long as the splitting $\Delta$ is non-zero, the vortex statistics is entirely non-universal. This differs from the conclusion of VS that vortices are fermions. The vortex statistics is fermionic only if $\Delta = 0$.

Thus, in the absence of an extra symmetry, which would guarantee the degeneracy of $\psi_+$ and $\psi_-$ vortices, the vortex statistics are non-universal. If such a symmetry is present, the statistics are fermionic.

\end{document}